\documentclass[aps,floatfix,twocolumn,prl,nofootinbib,superscriptaddress]{revtex4}
\usepackage{graphicx,bm,color}
\pdfoutput=1
\usepackage{array}
\usepackage{amsmath,amsfonts,amssymb,stackrel}
\usepackage{hyperref}
\hypersetup{colorlinks=true}
\hypersetup{urlcolor=blue}
\hypersetup{citecolor=black}
\hypersetup{menucolor=black}
\hypersetup{linkcolor=black}
\usepackage{cleveref}
\usepackage{longtable}
\usepackage{float}
\usepackage{pdfpages}
\usepackage[T1]{fontenc}
\usepackage[utf8]{inputenc}
\usepackage{bm}
\usepackage{tikz}
\usetikzlibrary{calc,patterns,decorations.pathmorphing,decorations.markings}
\PassOptionsToPackage{hyphens}{url}

\newcommand{\be}{\begin{equation}}
\newcommand{\ee}{\end{equation}}
\newcommand{\bea}{\begin{eqnarray}}
\newcommand{\eea}{\end{eqnarray}}

\newcommand{\meanv}[1]{\langle #1 \rangle}

\newcommand{\bb}[1]{\left( #1 \right)}

\newcommand{\ii}{\textrm{i}}
\newcommand{\dd}{\textrm{d}}
\newcommand{\eee}{\textrm{e}}

\newcommand{\kk}{\textbf{k}}

\tikzstyle{ressort}=[decorate,decoration={zigzag,pre length=0.0cm,post length=0.0cm,segment length=5, amplitude=0.1cm}]

\tikzset{->-/.style={decoration={
  markings,
  mark=at position .5 with {\arrow[scale=2,color=black]{>}}},postaction={decorate}}}
\tikzset{-<-/.style={decoration={
  markings,
  mark=at position .5 with {\arrow[scale=3,color=black]{<}}},postaction={decorate}}}

\tikzset{->>-/.style={decoration={
  markings,
  mark=at position .5 with {\arrow{>>}}},postaction={decorate}}}
\tikzset{-<<-/.style={decoration={
  markings,
  mark=at position .5 with {\arrow{<<}}},postaction={decorate}}}

\tikzset{phantom->-/.style={decoration={
  markings,
  mark=at position .5 with {\arrow[scale=2]{>}}},postaction={decorate}}}

\tikzset{serpent/.style={decoration={snake},postaction={decorate}}}

\begin{document}
\title{Interaction quenches in nonzero temperature fermionic condensates}

\author{H. Kurkjian}
\affiliation{Laboratoire de Physique Théorique,
Université de Toulouse, CNRS, UPS, 31400, Toulouse, France}
\author{V. E. Colussi}
\affiliation{Infleqtion, Inc., 3030 Sterling Circle, Boulder, CO 80301, USA}
\affiliation{Pitaevskii BEC Center, CNR-INO and Dipartimento di Fisica, Università di Trento, 38123 Trento, Italy}
\author{P. Dyke}
\affiliation{Optical Sciences Centre, ARC Centre of Excellence in Future Low-Energy Electronics Technologies, Swinburne University of Technology, Melbourne 3122, Australia}
\author{C. Vale}
\affiliation{Optical Sciences Centre, ARC Centre of Excellence in Future Low-Energy Electronics Technologies, Swinburne University of Technology, Melbourne 3122, Australia}
\author{S. Musolino}
\affiliation{Universit\'e C\^ote d'Azur, CNRS, Institut de Physique de Nice, 06200 Nice, France}
\affiliation{}

\begin{abstract}
We revisit the study of amplitude oscillations in a pair condensate of fermions after an interaction quench, 
and generalize it to nonzero temperature. {For small variations of the order parameter},  we show that the energy transfer during the quench
determines both the asymptotic pseudo-equilibrated value of the order parameter
and the magnitude of the oscillations, after multiplication by, respectively, the static response of the order parameter
and spectral weight of the pair-breaking threshold. Since the energy transferred to the {condensed pairs}
decreases with temperature as the {superfluid contact}, the oscillations eventually disappear at the critical temperature. 
{For deeper quenches}, we generalize the regimes of persistent oscillations and monotonic decay to nonzero temperatures, and explain 
how they become more abrupt and are more easily entered at high temperatures when the ratio of the initial to final gap either diverges,
when quenching towards the normal phase, or tends to zero, when quenching towards the superfluid phase.  Our results 
are directly relevant for existing and future experiments on the non-equilibrium evolution of Fermi superfluids near the phase transition.
\end{abstract}
\maketitle

\textit{Introduction.}---Fermionic condensates, unlike most of their bosonic counterparts,
are made of composite objects, known as Cooper pairs. This internal structure
implies more degrees of freedom beyond the usual sound waves found in bosonic systems \cite{PitaevskiiStringari}.
At the individual level, single Cooper pairs can break into two 
unpaired fermions, which leads to a gapped spectrum of fermionic quasiparticles \cite{Zwerger2009}. 
At the many-body level, whole wavepackets of quasiparticles can be excited 
for example by tuning the interparticle interaction strength \cite{Stringari2012}. This causes 
the amplitude of the order parameter to oscillate in a characteristic way \cite{Kogan1973}, with a frequency and damping
determined by the spectral distribution of the wavepacket. 

In contrast with the typical picture of amplitude or Higgs modes relying
on a single complex bosonic field \cite{Varma2015} in a Mexican hat potential, 
amplitude oscillations in a fermionic condensate are an intrinsically many-body 
effect, emerging only from the superposition of individual quasiparticle vibrations \cite{Schmid1968,Kogan1973,Gurarie2009,Foster2015}.
Still, for spatially-dependent and weak perturbations of the interaction strength, the evolution of the excited quasiparticle wavepacket 
can be summarized by a single pole of the order-parameter response function, 
such that the oscillations can be interpreted as a damped collective mode \cite{Popov1976,higgs}.

The case of homogeneous (zero-momentum) perturbations is more subtle: one can no longer identify
a pole in the order-parameter response function, such that the collective mode disappears.
There remains however a non-Lorentzian singularity in the spectral function, right at the threshold
energy for breaking Cooper pairs. In the time-domain, this converts into 
the famous power-law decaying oscillations of the order parameter \cite{Kogan1973}. 
The density of quasiparticle states available around the pair-breaking threshold changes
depending on whether the gapped fermionic spectrum has its minimum at zero or nonzero momentum, corresponding respectively
to the Bose-Einstein Condensate (BEC) or Bardeen-Cooper-Schrieffer (BCS) regimes. The lower density of states in the BEC regime
makes the damping
exponent increase to 3/2, compared to 1/2 in the BCS regime \cite{Gurarie2009}.

This remarkable collective effect has recently been the center of much
experimental attention, both with ultracold fermionic atoms \cite{Koehl2018,
swinburne}, superconductors \cite{Shimano2013} and cavity QED simulators \cite{Rey2021,Thompson2023}. The observations in those
experiments have revealed some important limits in our theoretical understanding of the oscillations.
Previous studies \cite{Gurarie2009,Foster2015}
have been restricted to zero temperature, whereas experimentally the oscillations 
have been recorded from low temperature to the vicinity of the phase transition. 
Additionally, important observables \cite{swinburne}, such 
as the oscillation amplitude, or the asymptotic limit of the order parameter, have not yet been fully understood.

Here, we show that oscillations of the order parameter {for small interaction quenches in the regime of linear response}
have the same form at zero and nonzero temperature: the power-law damping
retains the same exponent, and the oscillation frequency $2\Delta$ simply decreases
with temperature as the gap $\Delta$.
However, the presence of thermally excited quasiparticles before the quench
limits the variation of the order parameter, which, in contrast to the zero-temperature case,
no longer tends at long time to its value expected following an adiabatic change of the interaction strength.
We interpret the magnitude of the oscillations as the product
of the spectral weight of the pair-breaking threshold with the energy change during the quench,
itself related to the change in the scattering length through the contact.

We also argue that nonlinear effects increase near the critical temperature since the ratio of the initial
to final equilibrium gap $\Delta_i/\Delta_f$ either diverges or tends to zero when the depth
of the interaction quench is kept fixed. The regime of power-law damped oscillations is thus hidden
by the nonlinear regimes of persistent oscillations (regime III of Ref.~\cite{Foster2015}) or overdamped 
evolution (regime I), and the evolution in those two regimes becomes more abrupt compared to low temperatures.

\textit{Model.}---We consider a balanced two-component Fermi gas trapped in a three-dimensional volume $V$ at temperature $T=1/\beta$
(we use $\hbar=k_B=1$ throughout this Letter), with contact interactions between 
$\uparrow$ and $\downarrow$ components. The density $\rho$ of the gas fixes 
the Fermi wave number $k_F=(3\pi^2\rho)^{1/3}$, and the bare coupling constant $g$ is renormalized \cite{Zwerger2012}
to yield the appropriate $s$-wave scattering length $a$. In the mean-field approximation, the homogeneous system evolves according
to the time-dependent BCS equations \cite{Ripka1985}
\bea
\ii\partial_t c_\kk &=& (k^2/m) c_\kk+\Delta(1-2n_\kk), \label{HFB1}\\
\ii\partial_t n_\kk &=& \Delta c_\kk^* -\Delta^* c_\kk, \label{HFB2}
\eea
where $m$ is the atomic mass, 
$n_\kk=\meanv{\hat a_{\kk\uparrow}^\dagger \hat a_{\kk\uparrow}}=
\meanv{\hat a_{\kk\downarrow}^\dagger \hat a_{\kk\downarrow}}$ is
the momentum distribution, $c_\kk=\meanv{\hat a_{-\kk\downarrow} \hat a_{\kk\uparrow}}$ 
the pairing wavefunction, and $\Delta={g}\int \dd^3 k\, c_\kk/(2\pi)^3$ the order parameter.

Before the quench, the gas is at equilibrium at temperature $T_i$, chemical potential
$\mu_i$ and scattering length $a_i$. This corresponds
to the static solution the BCS equations, that is, the usual BCS thermal state with
$n_{\kk,i}=[1-(1-2F_\kk){\xi_\kk}/{\epsilon_\kk}]/2$ and $c_{\kk,i}=-(1-2F_\kk){\Delta_i}/{2\epsilon_\kk}$, 
in terms of the free-fermion and BCS dispersion relations, $\xi_\kk=k^2/2m-\mu_i$ and $\epsilon_\kk=\sqrt{\xi_\kk^2+\Delta_i^2}$,
and Fermi-Dirac distribution $F_\kk=1/(1+\eee^{\epsilon_\kk/T_i})$.
The abrupt variation of $a$ from $a_i$ to $a_f$ leaves the microscopic variables unchanged
($n_\kk(t=0^+)=n_{\kk,i}$ and similarly for $c_\kk$) but affects the coupling constant through 
\be
\frac{1}{g_f}-\frac{1}{g_i}=\bb{\frac{1}{a_f}-\frac{1}{a_i}}\frac{m}{4\pi}.
\label{lipp}
\ee
obtained via the Lippmann-Schwinger equation \cite{taylor2006scattering,Varenna}.

The initial kink in the order parameter then follows from the gap equation
\be
\Delta(t=0^+)-\Delta_i=\frac{g_f-g_i}{g_i} \Delta_i \label{changedelta}.
\ee
This kink corresponds to an energy variation that is
proportional to the extensive contact $C\equiv\dd (E/V)/\dd(1/a)$ \cite{Castin2012,Zhai2021,Tan2008,Tan2008_2,Tan2008_3,Braaten2008}:
\bea
\epsilon\equiv\frac{E_f-E_i}{V}=
- \frac{C}{4\pi m}\bb{\frac{1}{a_f}-\frac{1}{a_i}}.
 \label{injectedenergy}
\eea
{While Eq.~\eqref{injectedenergy} is valid in geneneral, 
the BCS approximation of the contact is $C_{\rm BCS}=m^2\Delta^2$.}
{This expression vanishes at the critical temperature as BCS theory 
approximates the normal phase by an ideal gas, and restricts the contact 
to the contribution of the condensed pairs.
For the general description of the gas, this is a rather crude approximation
in particular near the critical temperature,
but for the amplitude oscillations studied in this Letter, the superfluid contact
is precisely the important quantity\footnote{While the inverse coupling constants $1/g_i$ and $1/g_f$ diverge linearly with
a momentum cutoff, their difference does not according to Eq.~\eqref{lipp}. Thus
the injected energy (Eq.~\eqref{injectedenergy}) remains finite and nonzero, while the discontinuity
in $\Delta$ (Eq.~\eqref{changedelta}) vanishes. This is a consequence of the formally divergent
interaction energy $E_{\rm int}=\Delta^2/g$ according to BCS theory.}.}

\textit{Linear response.}---Shallow quenches are generally characterized by a small injected energy per particle
although this rule is brought into question later in this Letter.
In this weakly-excited regime, one can linearize the BCS system Eqs.~\eqref{HFB1}--\eqref{HFB2}
around the initial equilibrium state and solve using the Laplace
transformation \cite{Gurarie2009,higgslong}. With the initial condition \eqref{changedelta}, the phase of the order parameter
is not excited, and only its modulus evolves as:
\be
\Delta(t)=\Delta_{\infty}-\epsilon \int_{\omega_{\rm th}}^{+\infty}\frac{2\dd\omega}{\pi} \frac{\cos \omega t}{ \omega}\text{Im}f(\omega+\ii0^+).
\label{deltat}
\ee
This expression is composed of an asymptotic value $\Delta_{\infty}$ reached when $t\to+\infty$,
and a time-dependent, oscillatory part, written as the frequency integral of the order-parameter modulus-modulus response function
\be
f(z)=-\frac{M_{11}(z)}{\Delta_i (M_{11}(z)M_{22}(z)-M_{12}^2(z))}. \label{fz}
\ee
The linear response matrix $M_{ij}$ appearing here
is given by integrals over the internal degrees of freedom of the Cooper pairs
$M_{11}=(z^2-4\Delta_i^2)M_{22}/z^2=z^2\int \frac{d^3 k}{(2\pi)^3} \frac{1-2F(\epsilon_\kk)}{2\epsilon_\kk(z^2-4\epsilon_\kk^2)}$
and $M_{12}=M_{21}=\int \frac{d^3 k}{(2\pi)^3} \frac{z \xi_\kk [1-2F(\epsilon_\kk)]}{\epsilon_\kk(z^2-4\epsilon_\kk^2)}$. 

Quite intuitively, the final shift in $\Delta$, obtained when the oscillations have decayed,
is the product of the transferred energy and static modulus response $f(\omega=0)$:
\be
\Delta_{\infty}=\Delta_i+f(0)\epsilon. \label{deltainfty}
\ee
We identify here an important effect of temperature
on the post-quench dynamics. When $T_i=0$ the asymptotic gap
$\Delta_{\infty}$ matches the equilibrium gap $\Delta_f$ that would be reached
after an adiabatic change of the scattering length from $a_i$ to $a_f$. This is due to the
static modulus response saturating the injected energy $f(0)=\dd \Delta/\dd \epsilon$.
This is no longer true for $T_i>0$, and instead
\be
|\Delta_{\infty}-\Delta_i|<|\Delta_f-\Delta_i|.
\ee
In other words, the order parameter remains closer to its initial value than it would
under an adiabatic transformation, as a part of the injected energy
is absorbed by the thermally excited quasiparticles. 
Both at zero and nonzero initial temperature, the state reached asymptotically is not an equilibrium
state and, in particular, does not have a well-defined temperature.
To describe equilibration, the integrable BCS system should be replaced
by an ergodic model.
\begin{figure}
\includegraphics[width=0.5\textwidth]{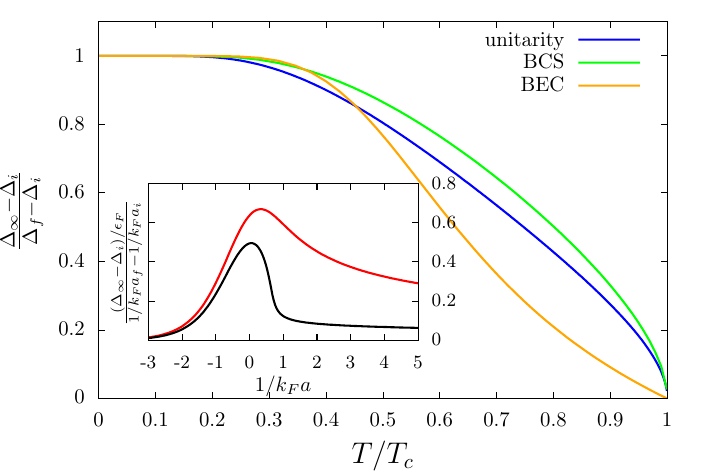}
\caption{\label{deltainfini} (Main panel) The asymptotic change of the order parameter $\Delta_{\infty}-\Delta_i$
measured relative to the change $\Delta_{f}-\Delta_i$ under an adiabatic evolution, as a function
of temperature at unitarity, in the BCS and BEC limits. At $T=0$, $\Delta_{\infty}=\Delta_f$ despite the non-adiabatic nature of the quench.
(Inset) The change $\Delta_{\infty}-\Delta_i$ (in units of $\epsilon_F$) relative to the change in the scattering length $1/k_F a_f-1/k_F a_i$.
throughout the BEC-BCS crossover at $T=0$ (red curve) and near $T_c$ ($T/T_c=0.99$, black curve). In both cases, a maximum is reached near unitarity.
Note that the change remains non zero (in units of $\epsilon_F$)
in the limit $T\to T_c$, which means the linear approximation breaks down (if the quench depth $1/k_F a_f-1/k_F a_i$ is kept independent of temperature).}
\end{figure}

Eq.~\eqref{deltainfty} provides a criterion for the validity of the linear regime.
For the deviation of the order parameter to remain small, it is necessary and sufficient that $|\Delta_{\infty}-\Delta_i|\ll \Delta_i$.
At low temperatures, $\Delta_i$ is comparable to the Fermi energy $\epsilon_F$, so this condition simply translates into
$|a_f-a_i|\ll a_i$, which is not a particularly demanding constraint, especially near unitarity ($1/|a|=0$).
Near $T_c$ however, $(\Delta_{\infty}-\Delta_i)/(1/k_F a_f-1/k_F a_i)$ is comparable to $\epsilon_F$
(as shown by the black curve in Fig.~\ref{deltainfini}) and hence much larger than $\Delta_i$.
This leads to a stricter condition $|a_f-a_i|\ll a_i\Delta_i/\epsilon_F$ for the validity of the linear approximation. 
For a quench depth $a_f-a_i$ fixed
independently of temperature, which corresponds to the experimentl scenario studied in \cite{swinburne}, 
this condition will {\it always} be violated when $T_i$ is sufficiently close to $T_c$.

\begin{figure}
\includegraphics[width=0.5\textwidth]{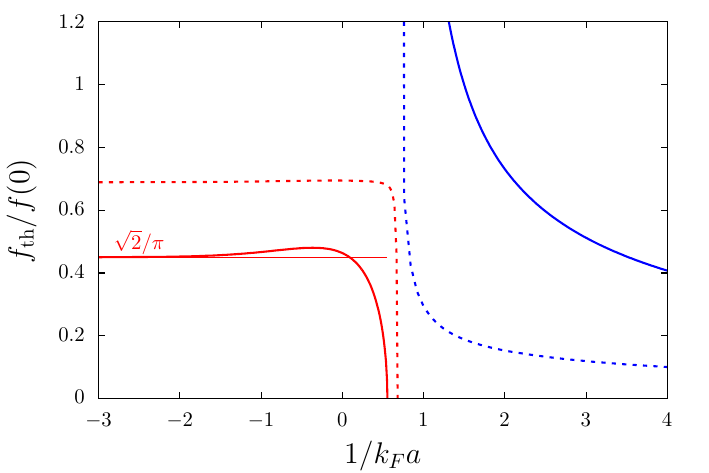}
\caption{\label{poidsspec} The spectral weight of the pair-breaking edge $f_{\rm th}$ (relative to the static response $f(0)$)
as a function of the interaction strength at $T=0$ (solid curves) and $T=0.999T_c$ (dashed curves). In red the BCS regime
where the edge exhibits a squareroot divergence (upper line of Eq.~\eqref{fth}), in blue the BEC regime where instead this edge
is a squareroot cancellation (lower line of Eq.~\eqref{fth}). Note that the boundary between those two regimes  depends weakly
on temperature.
} 
\end{figure}

We now turn to the time-evolution described by Eq.~\eqref{deltat}. The continuity of $\Delta(t)$ at $t=0$ is guaranteed by the 
sum-rule of the modulus-modulus response function: $\int_{+\infty+\ii0^+}^{-\infty+\ii0^+}{\dd z}f(z)/{2\ii\pi z}=0$. 
Then, at long times, the nature of the oscillations of $\Delta(t)$ depends on
the behavior of $f$ in the vicinity of the pair-breaking threshold
$\omega_{\rm th}$. In the BCS regime ($\mu_i>0$) and, irrespective of the temperature, the response function
has a squareroot divergence near $\omega_{\rm th}=2\Delta_i$. Conversely,
in the BEC regime ($\mu_i<0$) at all temperatures, the response function
is cancelled as a squareroot near the dimer-breaking threshold: 
\be
\text{Im}f(\omega+\ii 0^+)\underset{\omega\to\omega_{\rm th}}{\sim}
\begin{cases}
f_{\rm th}\sqrt{\frac{\omega_{\rm th}}{\omega-\omega_{\rm th}}} \text{ when } \mu_i>0 \\
f_{\rm th}\sqrt{\frac{\omega-\omega_{\rm th}}{\omega_{\rm th}}} \text{ when } \mu_i<0 
\end{cases}.
\label{fth}
\ee
After the frequency integration, these behaviours near $\omega_{\rm th}$ translate into power-law attenuated oscillations of $\Delta(t)$:
\be
\frac{\Delta(t)-\Delta_{\infty}}{\Delta_i-\Delta_\infty}\!\!\underset{t\to+\infty}{\sim}\!\!
\begin{cases}
{\frac{f_{\rm th}}{f(0)}}\sqrt{\frac{4}{\pi\omega_{\rm th} t}}{\cos\bb{\omega_{\rm th}t+\frac{\pi}{4}}}  \text{, } \mu_i>0 \\
\frac{f_{\rm th}}{f(0)}\frac{1}{\sqrt{\pi\omega_{\rm th}^3 t^3}}{\cos\bb{\omega_{\rm th}t+\frac{3\pi}{4}}} \text{, } \mu_i<0 
\end{cases}\label{GurarieForm}
\ee
The spectral weight $f_{\rm th}$ which characterizes the asymptotic
behaviors at the threshold is shown in Fig.~\ref{poidsspec} as a function of the interaction regime.
Comparing the zero-temperature case (solid curves) to the vicinity of $T_c$ (dashed curves), we observe 
a suppression of the relative weight $f_{\rm th}/f(0)$ on the BEC side but
an increase on the BCS side.
While this increase a priori favors the observability of the power-law damped oscillations,
we note that $f_{\rm th}/f(0)$ characterizes the amplitude of the signal only when scaled to the asymptotic change in $\Delta$,
see Eq.~\eqref{GurarieForm}. Scaled to the adiabatic variation $\Delta_f-\Delta_i$, the amplitude will
vanish as $\Delta_\infty-\Delta_i$ as shown by Fig.~\ref{deltainfini}.

\paragraph{Quenches in the nonlinear regime.}---The fact that the nonlinearity increases with temperature (as long as the quench depth $|a_i-a_f|$ is fixed) suggests extending our study to the nonlinear regime.
We do this numerically by simulating Eqs.~\eqref{HFB1}--\eqref{HFB2} on a fine momentum grid.

We recall the zero-temperature quench diagram of Ref.~\cite{Foster2015} (see Fig.~5 therein) that identified three qualitatively distinct regimes 
in the ($\Delta_i$, $\Delta_f$) plane. In regime I, there are no oscillations as $\Delta(t)$ is overdamped; this regime includes in particular the limit $\Delta_i\gg\Delta_f$. Regime II is the regime of power-law damped oscillations, which contains the linear regime on the diagonal $\Delta_i\simeq\Delta_f$. Finally, a regime III of undamped oscillations was identified around the limit $\Delta_f\gg\Delta_i$. 

We show now how regimes I and III generalize to nonzero temperatures and tend to hide regime II when the initial state approaches the critical point ($T_{i}\to T_{c,i}$) and the quench depth is fixed. 
For an initial state in the regime $|T-T_c|\ll T_c$, that is, $\Delta_i\ll\epsilon_F$, quenches in the direction of the superfluid phase end up in $\Delta_f\approx\epsilon_F\gg\Delta_i$, and therefore in regime III of persistent oscillations. Conversely, quenches towards the normal phase yield $\Delta_i\gg\Delta_f=0$, and thus fall into the overdamped regime I.

\begin{figure}
\includegraphics[width=0.495\textwidth]{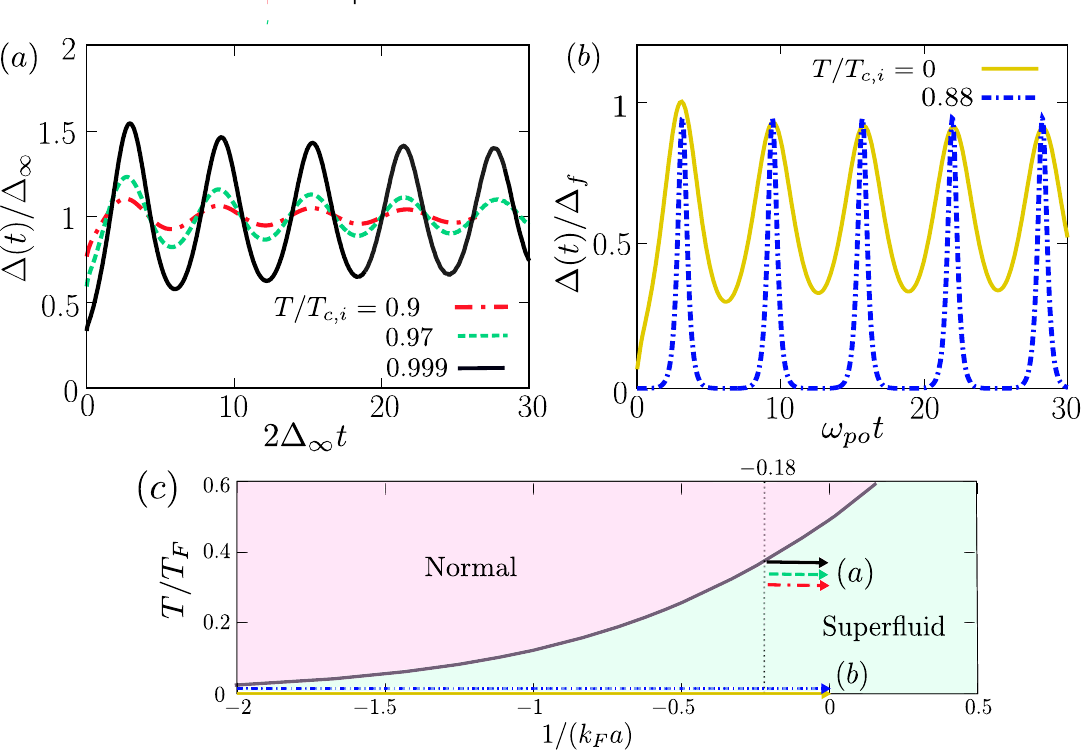} 
\caption{(a) The onset of regime III (persistent oscillations) in a quench from $1/(k_Fa_i)=-0.18$ to  $1/(k_Fa_f)=0$
when raising the initial temperature. (b) Effect of temperature on the persistent oscillations
for a quench from $1/(k_Fa_i)=-2$ to  $1/(k_Fa_f)=0$  belonging to regime III at all temperatures. Here, we find that the oscillation frequency $\omega_{po}$ is smaller than $2\Delta_f$. (c) Illustration of the quenches studied in (a) and (b) in the ($1/(k_F a)$, $T/T_F$) plot .} 
\label{fig:regimeIII} 
\end{figure}

In Fig.~\ref{fig:regimeIII}, we illustrate the onset of regime III at high temperatures when quenching in the direction of the superfluid phase. Going from the BCS side ($a_i<0$) to unitarity, with a quench depth sufficiently low to be in regime II at $T=0$, {as in Fig.~\ref{fig:regimeIII}(a)}, we notice an increase of the oscillation amplitude (scaled to $\Delta_\infty$), which precedes the appearance of persistent oscillations at temperatures close to $T_{c, i}$. The persistent oscillations also become much more abrupt than at low temperature, as illustrated by 
Fig.~\ref{fig:regimeIII}(b), where the quench depth is chosen to be in regime III already at $T=0$.

\begin{figure}
\includegraphics[width=0.495\textwidth]{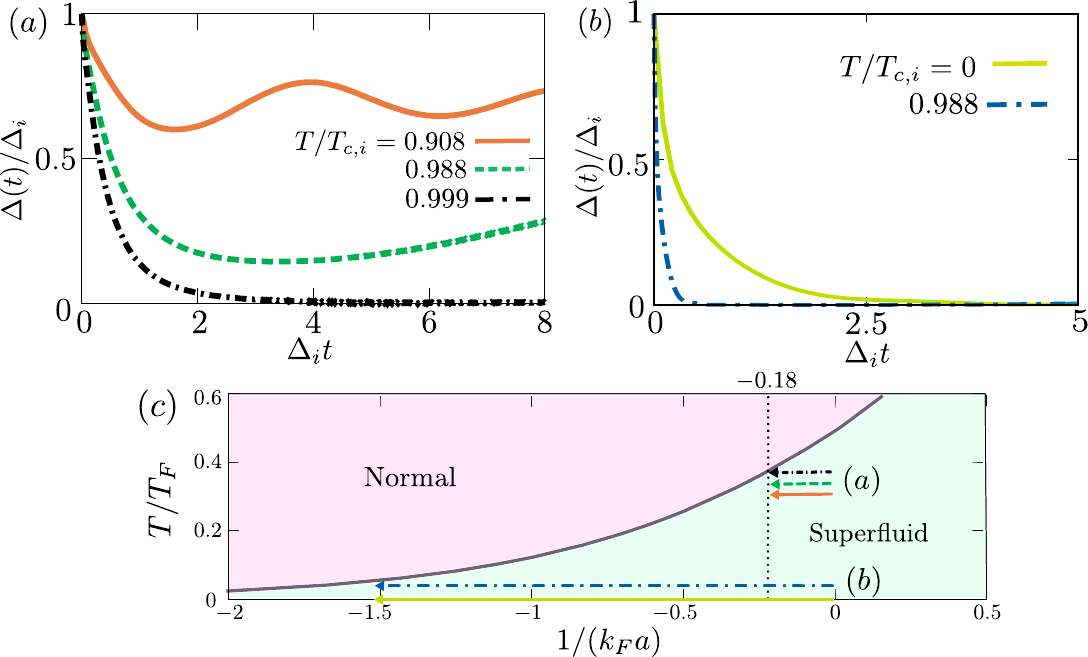} 
\caption{(a) The onset of regime I (overdamped evolution)  when raising the initial temperature of quenches from $1/(k_Fa_i)=0$ to  $1/(k_Fa_f)=-0.18$. (b) The decay of $\Delta(t)$ becomes more abrupt at higher temperature, as shown here for a quench from $1/(k_Fa_i)=0$ to  $1/(k_Fa_f)=-1.5$  belonging to regime I at all temperatures. (c) Illustration of the quenches studied in (a) and (b) in the ($1/(k_F a)$, $T/T_F$) plot.} 
\label{fig:regimeI} 
\end{figure}

In Fig.~\ref{fig:regimeI}, we consider the opposite case of quenches towards the normal phase with  $1/a_i = 0$ and $a_f$ on the BCS side. {As shown in Fig.~\ref{fig:regimeIII}(a)}, quenches sufficiently shallow to be in regime II at low temperatures undergo a gradual decrease of their asymptotic limit and oscillation frequency (both determined by $\Delta_\infty$) with temperature, up to a point where the order parameter tends to zero and no longer oscillates 
In Fig.~\ref{fig:regimeI}(a), this occurs at $T/T_c=0.999$, corresponding to $\Delta_f/\Delta_i \sim 2 \times 10^{-4}$. 
This threshold of regime I is reached at a lower temperature for larger quench depths. When the quench is sufficiently deep to be in regime I already at $T=0$, it remains in this regime at all temperatures, and the decay of $\Delta(t)$ becomes more abrupt as illustrated by Fig.~\ref{fig:regimeI}(b).

\paragraph{Conclusion.}---{We have studied amplitude oscillations in a nonzero temperature fermionic condensate within time-dependent BCS theory.
We showed how the magnitude of the oscillations and the asymptotic change of the order parameter $\Delta_\infty-\Delta_i$ are both proportional to the BCS contact.
The oscillations thus fade out as this contact vanishes at the phase transition.
While the oscillation frequency is predicted to vanish at $T_c$ with $\Delta$ on the BCS side,
it stays nonzero on the BEC side and coincides with the molecular binding energy $E_{\rm mol}=2|\mu|$.
The fact that time-dependent BCS theory does not correctly describe the normal 
phase of the interacting gas limits our description of what happens outside the superfluid phase, in during particular dynamical phase
transitions \cite{Vale2021}. Extending BCS theory to correctly describe the nonequilibrium evolution across the phase transition
would be a major achievement.
One can also imagine that amplitude oscillations of the pairing field still occur
in the normal phase when a pseudogap appears in the single-particle spectral density \cite{Dong2010}. 


\begin{acknowledgments}
We thank Denise Ahmed-Braun, Servaas Kokkelmans for
valuable discussions.  V.E.C. acknowledges financial support from the Provincia Autonoma di Trento, the Italian MIUR under the PRIN2017 projectCEnTraL and the National Science Foundation under Grant No. NSF PHY-1748958. S. M. acknowledges financial support from the ANR-21-CE47-0009 Quantum-SOPHA project.
\end{acknowledgments}

\bibliography{biblio}


\end{document}


\title{Supplementary material: Interaction quenches in nonzero temperature fermionic condensates}

\author{H. Kurkjian}
\author{V. E. Colussi}
\author{P. Dyke}
\author{C. Vale}
\author{S. Musolino}

\begin{abstract}
\end{abstract}

\makeatletter
\def\NAT@spacechar{}
\makeatother

\maketitle

\section{Linearisation of the HFB equations}

We give here additional details on our analytical solution of the BCS
equations \eqref{HFB1}--\eqref{HFB2} in the regime of weak perturbations,
leading to Eq.~\eqref{deltat}. We linearize the evolution
around the initial equilibrium state, rather than around a virtual final equilibrium state
as in Ref.~\cite{Gurarie2009}. At nonzero temperature, where $\Delta(t)$ never reaches
$\Delta_f$, this is far more intuitive.

We introduce the fluctuations $\delta c_\kk=c_\kk-c_{\kk,i}$, 
$\delta n_\kk=n_\kk-n_{\kk,i}$, and $\delta\Delta=\Delta-\Delta_i$ and linearize the BCS system \eqref{HFB1}--\eqref{HFB2}:
\bea
\!\!\!\!\!\!\!\ii\partial_t \delta c_\kk &=& {\color{black} 2\xi_\kk  \delta c_\kk -2\Delta_i\delta n_\kk}+{\color{black} \frac{\xi_\kk }{\epsilon_\kk}F(\epsilon_\kk)\delta\Delta} \label{dck}\\
\!\!\!\!\!\!\!\ii\partial_t \delta n_\kk &=& {\color{black} -\Delta_i(\delta c_\kk-\delta c_\kk^*)}-{\color{black} \frac{\Delta_i}{2\epsilon_\kk}F(\epsilon_\kk) (\delta\Delta-\delta\Delta^*)} \label{dnk}
\eea
Although the quench scenario corresponds to $\delta n_\kk=\delta c_\kk=0$ at $t=0^-$, 
we make so far no assumption on the initial state.
In this more general case, the fluctuation of $\Delta$ has a time-dependent part caused by the $\delta c'$s,
and a constant part $\delta\Delta_0=\frac{g_f-g_i}{V}\sum_\kk c_{\kk,i}$ caused by the quench on $g$:
\be
\delta\Delta(t)=\frac{g_i}{V}\sum_\kk \delta c_\kk(t) +\delta\Delta_0
\ee

In the spirit of Ref.~\cite{artmicro}, we now move to the quasiparticle basis:
\bea
\alpha_\kk^+ &=&
\frac{\xi_\kk}{\epsilon_\kk} (\delta c_\kk+\delta c_\kk^*)-\frac{2\Delta_i}{\epsilon_\kk}\delta n_\kk\notag\\
\alpha_\kk^- &=&
\delta c_\kk-\delta c_\kk^*\\
m_\kk &=& 
\!\!\!\frac{2\xi_\kk}{\epsilon_\kk} \delta n_\kk+\frac{\Delta_i}{\epsilon_\kk}(\delta c_\kk+\delta c_\kk^*)
\eea
so as to diagonalise the individual parts of Eqs.~\eqref{dck}--\eqref{dnk}:
\bea
\ii\partial_t \alpha_\kk^+ &=& {\color{black} 2\epsilon_\kk \alpha_\kk^- }+{\color{black}F(\epsilon_\kk)(\delta\Delta-\delta\Delta^*)} \label{alpk}\\
\ii\partial_t \alpha_\kk^- &=& {\color{black} 2\epsilon_\kk \alpha_\kk^+}+{\color{black} \frac{\xi_\kk}{\epsilon_\kk}F(\epsilon_\kk) (\delta\Delta+\delta\Delta^*)} \label{almk} \\
\ii\partial_t m_\kk &=& 0
\eea
In the quasiparticle basis, the fluctuations of $\Delta$ take the form:
\be
\delta\Delta = \delta\Delta_0 + \frac{g_i}{2}\int \frac{\dd^3 k}{(2\pi)^3} \bbcro{\alpha_\kk^-+\frac{\xi_\kk}{\epsilon_\kk} \alpha_\kk^+ + \frac{\Delta_i}{\epsilon_\kk}m_\kk} \label{deltap}
\ee
To solve the time-dependent system, we introduce the Laplace transform
of the variables:
\be
A_\kk^\pm(\omega)=\int_{0^-}^{+\infty} \eee^{\ii\omega t} \alpha_\kk^\pm(t) \dd t
\ee
and similarly for $M_\kk(\omega)$ and $\delta_\pm(\omega)$ the transform of $m_\kk(t)$ and $\delta\Delta(t)\pm\delta\Delta^*(t)$ respectively. 
This allows us to express the microscopic variables in terms of the fluctuations of $\Delta$,
\begin{widetext}
\bea
{\color{black} A_\kk^+(\omega)}&=& \frac{\ii\omega\alpha_\kk^+(0^-)+2\ii\epsilon_\kk\alpha_\kk^-(0^-)}{\omega^2-4\epsilon_\kk^2} +\frac{2\xi_\kk F(\epsilon_\kk)}{\omega^2-4\epsilon_\kk^2}{\color{black} \delta_+(\omega)}+ \frac{\omega  F(\epsilon_\kk)}{\omega^2-4\epsilon_\kk^2}{\color{black} \delta_-(\omega)} \\
{\color{black} A_\kk^-(\omega)}&=& \frac{\ii\omega\alpha_\kk^-(0^-)+2\ii\epsilon_\kk\alpha_\kk^+(0^-)}{\omega^2-4\epsilon_\kk^2}+\frac{2\epsilon_\kk F(\epsilon_\kk)}{\omega^2-4\epsilon_\kk^2} {\color{black} \delta_-(\omega)} +\frac{\omega \xi_\kk F(\epsilon_\kk)}{\epsilon_\kk(\omega^2-4\epsilon_\kk^2)}{\color{black} \delta_+(\omega)} \\
{\color{black}  M_\kk(\omega)}&=& \frac{\ii m_\kk(0^-)}{\omega}
\eea
\end{widetext}
and finally to eliminate them using the resummation Eq.~\eqref{deltap}, yielding a closed system of equations on $\delta_\pm$:
\be
M\begin{pmatrix} \delta-\delta^* \\ \delta+\delta^* \end{pmatrix}=-\ii\begin{pmatrix} S_- \\ S_+ \end{pmatrix}
\label{MSpm}
\ee
The fluctuation matrix $M$ is introduced below equation \eqref{fz} of the main text,
and the sums encoding the initial conditions on $\Delta$ are given by
\begin{multline}
\!\!\!\!\!\!S_+ \!=\!\int \frac{\dd^3\kk}{(2\pi)^3}\! \bbcro{\frac{\omega \xi_\kk\alpha_\kk^+(0^-)+2\xi_\kk\epsilon_\kk\alpha_\kk^-(0^-)}{\epsilon_\kk(\omega^2-4\epsilon_\kk^2)} \! + \!  \frac{\Delta_i}{\epsilon_\kk}\frac{ m_\kk(0^-)}{\omega}} \label{Sm} \\
+\frac{\delta\Delta_0+\delta\Delta_0^*}{g_i \omega}
\end{multline}
\begin{multline}
\!\!\!\!S_- = \int  \frac{\dd^3\kk}{(2\pi)^3} \frac{\omega\alpha_\kk^-(0^-)+2\epsilon_\kk\alpha_\kk^+(0^-)}{\omega^2-4\epsilon_\kk^2}  \label{Sp}
+\frac{\delta\Delta_0-\delta\Delta_0^*}{g_i \omega}
\end{multline}
Inverting Eq.~\eqref{MSpm} and switching back to the time domain [using the inverse Laplace transformation $f(t)=-\frac{1}{2\pi}\int_{+\infty+\ii\eta}^{-\infty+\ii\eta}$ $\dd z\eee^{-\ii z t}F(z)$],
yields, for the time-evolution of $\Delta$:
\be
\begin{pmatrix}
\delta\Delta(t)-\delta\Delta^*(t)
\\
\delta\Delta(t)+\delta\Delta^*(t)
\end{pmatrix}=-\int_{+\infty+\ii\eta}^{-\infty+\ii\eta}\frac{\dd z}{2\ii\pi} \eee^{-\ii zt}  M^{-1} \begin{pmatrix} S_-\\ S_+
\end{pmatrix}
\label{Deltat}
\ee
We now input the initial condition corresponding to the interaction quench, that is
(as explained above Eq.~\eqref{lipp} of the main text) $\alpha_\kk^\pm=m_\kk=0$,
and $\delta\Delta_0=\epsilon g_i/\Delta_i$, which converts into $S_-=0$ and $S_+=2\epsilon/\omega\Delta_i$.
Finally, we derive Eq.~\eqref{deltat} of the main text by 
closing the integration contour in Eq.~\eqref{Deltat} 
around the branch cuts of $M$ (see Fig.~1 in Ref.~\cite{Gurarie2009}),
and by remarking that the modulus-modulus response function (Eq.~\eqref{fz}) is related to $M$
by $f(z)=-(M^{-1})_{22}/\Delta_i$.

\section{Calculation of $M$}

We numerically evaluate the elements of $M$ at nonzero temperature using the Kramers-Kronig relation
\bea
\frac{M_{11}}{\omega_0^2}=\frac{M_{22}}{\omega_0^2-4\Delta^2}&=&\int_{-\infty}^{+\infty}\frac{\rho_{f}(\omega)\dd\omega}{\omega_0-\omega}\label{M11}\\
{M_{12}}&=&\int_{-\infty}^{+\infty}\frac{\rho_{g}(\omega)\dd\omega}{\omega_0-\omega}\label{M12}
\eea
In dimensionless units ($\check\rho=(2\pi)^3\Delta\rho/k_\Delta^3$, $\check\omega=\omega/2\Delta$ $\check\beta=\beta\Delta$), we have 
\bea
\check\rho_{f}&=&\frac{\pi}{2\check\omega}\frac{\text{th}(\check\beta\check\omega)}{\sqrt{\check\omega^2-1}}\bb{k_1(\check\omega)+k_2(\check\omega)}\\
\check\rho_{g}&=&\frac{\pi}{2}{\text{th}(\check\beta\check\omega)}\bb{k_1(\check\omega)-k_2(\check\omega)}
\eea
with
\begin{widetext}
\be
\begin{cases}
k_1(\omega)=\Theta(\check\omega-1)\sqrt{\check\mu+\sqrt{\check\omega^2-1}} \\
k_2(\omega)=\Theta(\check\omega-1)\Theta(\sqrt{1+\check\mu^2}-\check\omega)\sqrt{\check\mu-\sqrt{\check\omega^2-1}} 
\end{cases}
\text{if }\check\mu>0\ 
\begin{cases}
k_1(\omega)=\Theta(\check\omega-\sqrt{1+\check\mu^2})\sqrt{\check\mu+\sqrt{\check\omega^2-1}} \\
k_2(\omega)=0
\end{cases}
\text{if }\check\mu<0
\ee
\end{widetext}
The divergence of the integrals \eqref{M11}--\eqref{M12} in $\omega=\omega_0$ is easily compensated by adding/subtracting
$\rho(\omega_0)$. Still,
the evaluation can become difficult when $\check\omega_0$ is close to the angular points $1$ or $\sqrt{1+\check\mu^2}$. To overcome this, we subtract the nearly diverging behavior of the integrand. (For here on, all quantities are implicitly dimensionless).
\begin{widetext}
\begin{itemize}
\item When $\omega_0\to1^+$, the integrand nearly diverges around $\omega\to 1^+$. We set $e=(\omega-1)/(\omega_0-1)$ and derive:
\bea
\frac{\rho_{f}(\omega)-\rho_{f}(\omega_0)}{\omega_0-\omega}&=&\frac{\pi\text{th}\beta}{\sqrt{2}}\frac{\sqrt{\mu}}{(\omega_0-1)^{3/2}}\frac{1}{\sqrt{e}+e} +O(\omega_0-1)^{-1/2}\\
\frac{\rho_{g}(\omega)-\rho_{g}(\omega_0)}{\omega_0-\omega}&=&O(\omega_0-1)^{-1/2}
\eea
\item When $\omega_0\to \omega_3^\pm$, the integrand nearly diverges around $\omega\to \omega_3^\mp$. We set $e=(\omega-\omega_3)/(\omega_3-\omega_0)$ and derive:
\bea
\frac{\rho_{f}}{\omega_0-\omega}&=&\frac{\pi\text{th}(\beta\omega_3)}{\sqrt{2}}\frac{1}{\omega_3\sqrt{\mu}}\frac{1}{\omega_0-\omega_3}\frac{1}{1+e} +O(1)\\
\frac{\rho_{g}}{\omega_0-\omega}&=&\frac{\pi\text{th}(\beta\omega_3)}{\sqrt{2}}\frac{\sqrt{\mu}}{\omega_0-\omega_3}\frac{1}{1+e} +O(1)
\eea
\end{itemize}
\end{widetext}
\bibliography{../biblio}
